\documentclass[twocolumn,showpacs,preprintnumbers,amsmath,amssymb]{revtex4}

\usepackage{graphicx}
\usepackage{dcolumn}
\usepackage{bm}

\begin{document}

\title{Dissociation of relativistic nuclei in \\peripheral interactions in nuclear track
 emulsion}
\thanks{The work was supported by the Russian Foundation for Basic Research (Grants nos. 96-1596423,
 02-02-164-12a, 03-02-16134, 03-02-17079, 04-02-16593, 04-02-17151),
 the Agency for Science of the Ministry for Education of the Slovak Republic and the Slovak Academy
 of Sciences (Grants VEGA 1/9036/02 and 1/2007/05) and  Grants from the JINR Plenipotentiaries
 of Bulgaria, Czech Republic, Slovak Republic, and Romania  during years 2002-5.}

\author{D.~A.~Artemenkov}
   \affiliation{Joint Insitute for Nuclear Research, Dubna, Russia}
 \
\author{G.~I.~Orlova}
   \affiliation{Lebedev Institute of Physics, Russian Academy of Sciences, Moscow, Russia} 
\  
 \author{P.~I.~Zarubin}
   \affiliation{Joint Insitute for Nuclear Research, Dubna, Russia}

\date{\today}

\begin{abstract}
 Possibilities of the nuclear emulsion technique for the study of the systems
 of several relativistic fragments produced in the peripheral interactions of relativistic
 nuclei are discussed. The interactions of the $^{10}$B and $^9$Be nuclei in emulsion are
 taken as an example to show the manifestation of the cluster degrees of freedom in
 relativistic fragmentation. For the case of the relativistic $^9$Be nucleus dissociation it is shown that exact
 angular measurements play a crucial role in the restoration of the excitation spectrum of
 the alpha particle fragments. The energy
 calibration of the angular measurements by the $^9$Be nucleus enables one to conclude
 reliably about the features of internal velocity distributions in more complicated systems
 of relativistic $\alpha$ particles. \par  

\end{abstract}
\maketitle
\section{\label{sec:level1}Introduction}
\indent The peripheral collisions of nuclei proceeding at energy above 1~A~GeV are
 collisions of a special type in which the breakup of the primary nuclei is provoked
 by electromagnetic and diffraction interactions, as well as by nucleon collisions for
 a minimal overlap of nuclear densities. Nuclear track emulsions exposed to beams of relativistic
 nuclei make it possible to obtain  an information about the charged products of such
 collisions which is unique as concerns details of observation of particle tracks and the
 accuracy of their spatial metrology \cite{Friedlander83,Baroni90,Baroni92}.\par
 \begin{figure*}
\footnotesize
 \centerline{\begin{tabular}{@{}cc@{}}
 \includegraphics[height=1.5in]{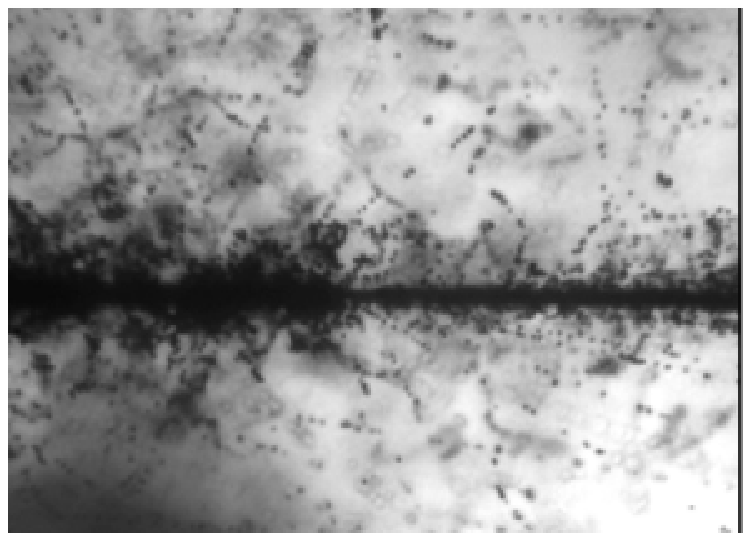} &
 \includegraphics[height=1.5in]{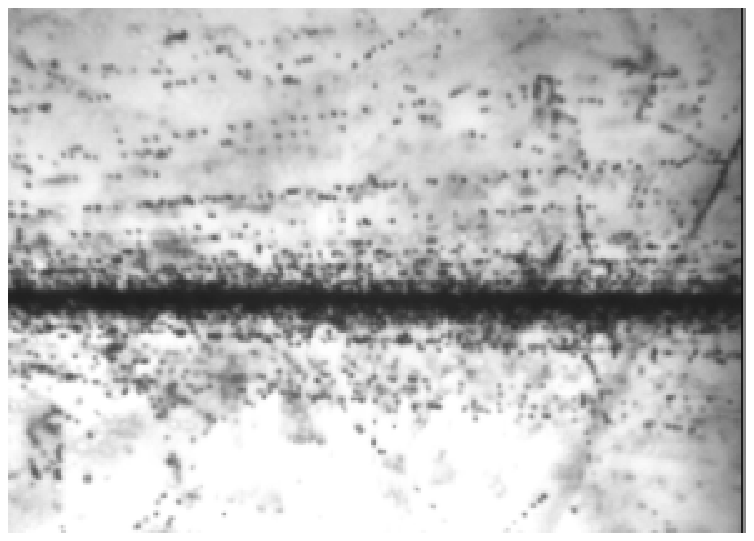}\\
 Shot 1 & Shot 2 \\[8bp]
 \includegraphics[height=1.5in]{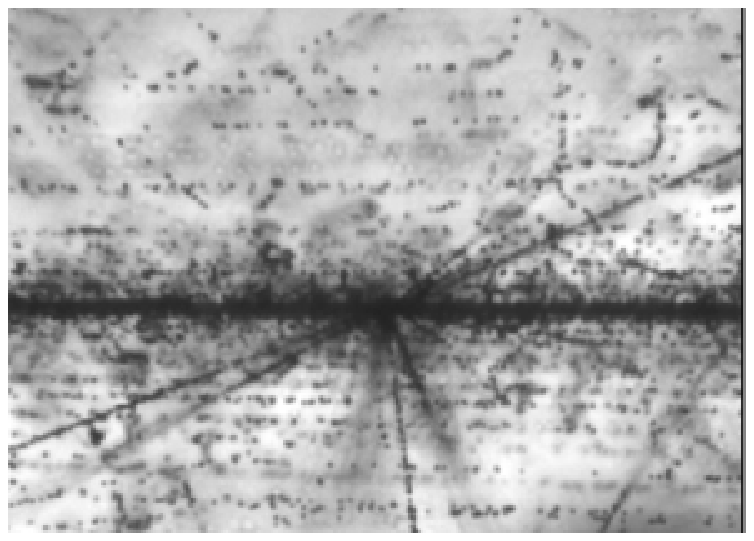} &
 \includegraphics[height=1.5in]{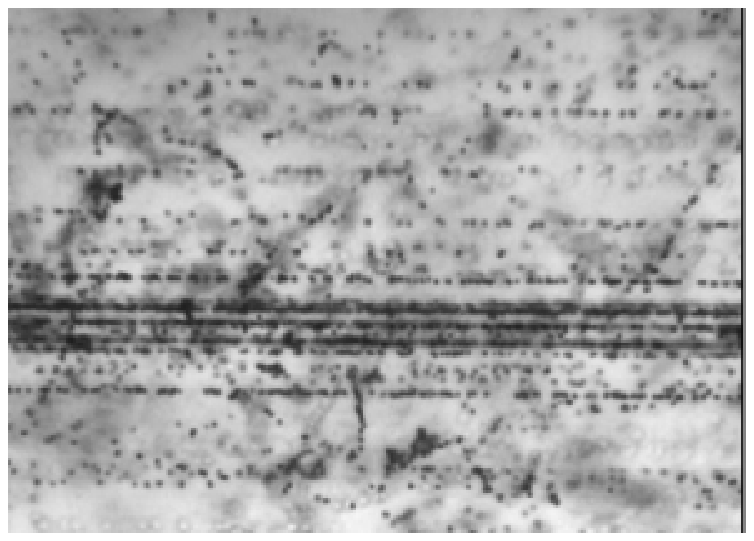}\\
 Shot 3 & Shot 4
 \end{tabular}}
\caption{\label{fig:1}Subsequently photoed event of peripheral interaction of a 158~A~GeV $^{207}$Pb nucleus in a nuclear track emulsion  
in $\approx$100$\times$100 $\mu$m$^2$ viewing fields:
 primary nucleus track and interaction vertex followed by projectile fragment jet (Shot~1); 
jet core with apparent tracks of singly and doubly charged particles (Shot~2); jet core with a secondary interaction star (Shot~3);
completely recognizable jet core (Shot~4, 3 cm  distance from the vertex).}
\end{figure*}    

\section{\label{sec:level2}Relativistic fragmentation}
\indent In peripheral interactions, nuclei are given an excitation spectrum near the energy
 dissociation thresholds. In the kinematical region of fragmentation of a relativistic
 nucleus, there arise systems consisting of nuclear fragments whose total charge is close
 to the parent nucleus charge. The opening angle of the relativistic fragmentation cone
 is defined by the Fermi nucleon motion. Thus, the fragments find themselves on the periphery
 of the particle rapidity distribution which is obtained by summing over all the channels of
 the reaction in question.\par
 \indent The values of the fragment momenta normalized to the mass numbers are distributed
 about the normalized momentum of the primary nucleus with few-percent dispersion.
 Therefore the distribution of the velocities of fragments in their c.m.s. must be a
 non-relativistic one. In accordance with the established pattern of the nuclear limiting
 fragmentation the probabilities of population of the fragment final states reveal a very
 high degree of universality. They are found to be weakly dependent on the initial energy
 and target-nucleus properties.\par
\indent The interactions of the above-mentioned type can serve as a \lq\lq laboratory\rq\rq
 ~for the generation of non-relativistic ensembles of several lighter nuclei. The term \lq\lq peripheral\rq\rq
 ~does not reflect in full measure dramatic changes which occur at the microscopic level.
 The dissociation degree of a nucleus can reach its total destruction into separate nucleons
 and lightest nuclei   having no excited states, that is, $^{2,3}$H and $^{3,4}$He nuclei.
 A relative intensity of their production permits one to reveal the importance of different
 cluster degrees of freedom.\par
\indent For the experimental study of multi-particle systems, the choice of those of them
 which result from the dissociation of a relativistic projectile and not from the dissociation
 of the target-nucleus has special methodical advantages. Owing to the kinematical collimation
 and absence of the detection threshold, relativistic fragments can completely be observed
 in a small solid angle and the distortions due to energy ionization losses in the detector
 matter are minimal.\par
\indent When selecting events with the dissociation of a projectile into the fragmentation
 cone, the non-relativistic fragments are either absent (\lq\lq white\rq\rq stars), or their number is insignificantly
 small. These fragments are emitted over all the solid angle, therefore, their fraction in
 the relativistic fragmentation angular cone is negligible. The target-fragments have
 non-relativistic momenta which allow one to distinguish them from the projectile fragments
 in this cone.\par
\indent Of course, in the relativistic approach to the fragmentation study, there also arise
 its own methodical troubles. For a primary nucleus with charge Z, it is very desirable to
 provide the detection up to singly-charged particles. The ionization produced by all
 fragments can be reduced down to a factor Z, while the ionization per one track - to a
 factor of Z$^2$ as compared with that from the primary nucleus. Therefore the experimental
 method should provide the widest detection range taking the Z$^2$ value into account.\par   

\section{\label{sec:level3}Capabilities of nuclear track emulsion}
\indent To reconstruct an event the full kinematical information about the secondary particles
 in the relativistic fragmentation cone is needed which, i. e. allow one to calculate the
 invariant mass of the system. The accuracy of its estimation drastically depends on the
 accuracy of the track angular resolution. Hence, to provide the best angular resolution the
 detection of fragments with the best spatial resolution is needed.\par
\indent At the initial stage of investigations, the nuclear emulsion method well satisfies
 these requirements. The major task of it is to search for reliable proofs of the existence
 of different fragmentation channels for a statistical provision at the level of dozens of
 events. Emulsions provide a record spatial resolution (about 0.5~$\mu$m)
 which makes it possible to separate the charged particle tracks in the three-dimensional
 image of an event within one layer thickness (600~$\mu$m), as well as ensure a high accuracy
 of measurement of the angles. The emulsion technique allows one to measure the particle charges,
 starting with the single-charged particles up to the highest-charged ones, by combining the
 ionization means (counting of the number of breaks and the number of $\delta$ electrons per
 track length unit). The tracks of relativistic H and He nuclei are distinguished
 by vision. In the peripheral fragmentation of a light nucleus its charge can often be
 established by the charge topology of relativistic fragments. A collection of appropriate reaction 
images can be found in \cite{Bradnova04} and in the BECQUEREL project web site \cite{web05}. Multiple scattering measurements
 on the light fragment tracks enable one to separate the $^{2,3}$H and $^{3,4}$He isotopes.\par
 
\indent A vivid illustration of these assertions is the microphotograph of the event of a total
 disintegration of Pb nucleus of energy 158 A GeV in its peripheral interaction with an
 emulsion nucleus~(see Fig.\ref{fig:1}). The exposure was performed in beams from the SPS
 accelerator (CERN) in the framework of the EMU collaboration. 
Experiment details can be found in \cite{EMU97,Adamovich99}.
 Shot 1 shows the primary nucleus track which is surrounded by a dense cloud of
 $\delta$ electrons. On Shot 2 the interaction vertex looks like a stepped lowering of
 the ionization density in which there are no tracks from the target-nucleus fragmentation.
 Shot 3 shows a gradual separation of the tracks of singly and doubly charged particles from
 the shower trunk. At a given energy the He nucleus emission angles are restricted to a
  0.1$^{\circ}$ value. A total separation of tracks is seen on Shot 4 corresponding to a
 distance of about 3 cm from the vertex. The observer does not see in this event an intense
 flux of dozens of relativistic neutrons that have not to be able to bind the lightest nuclei.
 The image of an event in emulsion is created by microscopic crystals of 1 $\mu$m in thick, i.e.
 the latter are larger than the real sizes of nuclear fragments by about 9 orders of magnitude.
 Nevertheless this image reproduces rather well details of a \lq\lq catastrophe\rq\rq ~occurred
 at the micro-world scale.\par
\indent The events of a total disintegration make up a small fraction of all the variety of
 the final states of heavy nuclei which embraces pairing fission, formation of single
 fragments accompanied by a great number of the lightest nuclei, formation of groups of
 light nuclei \cite{Friedlander83,Jain84,Adamovich97,Adamovich98,Cherry98}. 
The excitation transferred to the nucleus is, to a large extent, defined by
 the energy threshold of the final-state mass. It grows with increasing fragment multiplicity.
 In this sense, the charge topology of the final state already defines the excitation.
 In a complicated process of the energy distribution over the multiplicity of the degrees
 of freedom, nuclear fragments go onto   the mass surface and get some possibility to realize
 the Coulomb energy of mutual repulsion into the kinetic energy of each fragment. Some kind
 of a Coulomb \lq\lq explosion\rq\rq of a nucleus occurs.\par           
\indent The example of a total disintegration of a Pb nucleus may be interpreted as an event
 of the phase transition of nuclear matter from the state of quantum liquid to the state of
 quantum dilute gas of nucleons and the lightest nuclei. The metrology of such events is
 laborious and requires a high level of skill. Nevertheless such events are of an undoubted
 scientific interest, therefore their accumulation continues by the BECQUEREL collaboration \cite{web05}.
 The light nucleus fragmentation can be considered as a component of the heavy nucleus
 fragmentation picture. In what follows, some examples are given to consider the role of the
 cluster degrees of freedom in the light nucleus fragmentation, as well as the energy scale
 of inter-cluster interactions.\par 
\section{\label{sec:level4}Clustering in light nuclei}
\indent The charge topology of the relativistic fragmentation of N, O, Ne, Mg and Si nuclei
 in peripheral interactions in emulsion is presented 
in \cite{El-Naghy88,Baroni90,Baroni92,Jilany04,Andreeva05}. A special feature of the
 excitation increase in this group of nuclei consists in the growth of the multiplicity of
 the He and H nuclei with decreasing charge of the only fragment with Z$>$3. In light nuclei
 the pairing splitting channel is practically suppressed.\par
 
\indent More specific correlation studies were performed 
for the leading fragmentation channels like
$^{12}$C$\rightarrow$3$\alpha$ \cite{Belaga95}, 
 $^{16}$O$\rightarrow$4$\alpha$ \cite{Andreeva96,Glagolev01}, 
$^{6}$Li$\rightarrow$d$\alpha$ \cite{Lepekhin98,Adamovich99},
$^{7}$Li$\rightarrow$t$\alpha$ \cite{AdamovichPh04},
$^{10}$B$\rightarrow$d$\alpha\alpha$ \cite{Adamovich04}, 
and $^{7}$Be$\rightarrow^{3}$He$\alpha$ \cite{BradnovaA04}.
In addition to the $\alpha$ clustering, a clustering  of nucleons in the form of  deuterons in $^{6}$Li and $^{10}$B decays,
 as well as of tritons in $^{7}$Li decays has been revealed. Besides, the multiparticle dissociation is found to be important
  for these nuclei. Emulsions exposed to relativistic $^{14}$N and $^{11}$B isotopes are being analyzed with the aim to study 
clustering of these types. A $^{3}$He clustering in $^{7}$Be relativistic excitation is demostrated  
 \cite{BradnovaA04,Andreeva05}. The next round of research, as to whether 
this kind of nuclear clustering is revealed in light neutron-deficient 
nuclei like $^{8}$B and $^{9,10,11}$C is in progress now at the JINR Nuclotron \cite{Malakhov04}.\par
\indent The decay of the excited states in the Be, B and C isotopes is of a clearly expressed
 $\alpha$ cluster character. In overcoming the mass threshold of a reaction their dissociation
 proceeds through the formation of an unstable $^8$Be nucleus in the ground and excited states.
 Among the reaction channels, 3-particle decays into He and H nuclei are dominant.
 Fragments with Z$>$3 do not play a crucial role.\par
\indent As an important application, this conclusion can affect the problems of cosmic-ray
 physics related to the element abundance in the region of a Li-Be \lq\lq gap\rq\rq .
 The fundamental problem of a  Li, Be, and B abundance in galactical cosmic rays 
 as compared with their abundance in the matter of the Solar system has not
 been solved yet. This pattern points out that the main chain of subsequent splitting of
 nuclei, when they are propagate in interstellar H and He gases, passes over the production of
 the Li, Be and B nuclei. This fact greatly stimulates the interest in the search for the
 sources of origin of the mentioned group of nuclei, especially the $^{6,7}$Li isotopes.\par
\begin{figure*}
\footnotesize
 \centerline{\begin{tabular}{@{}cc}
 \includegraphics[height=0.75 in]{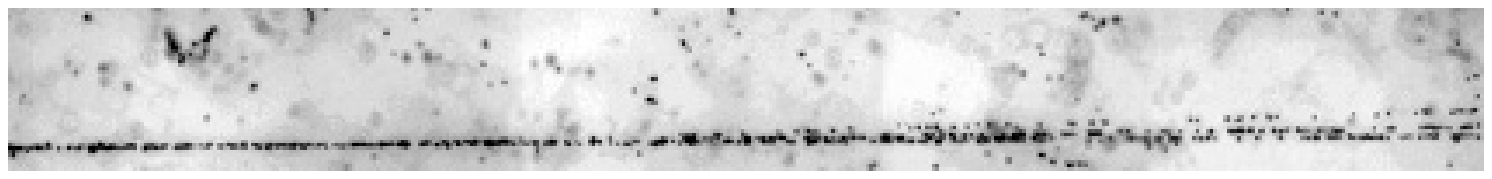}\\
  Shot~1  \\[8bp]
 \includegraphics[height=0.74 in]{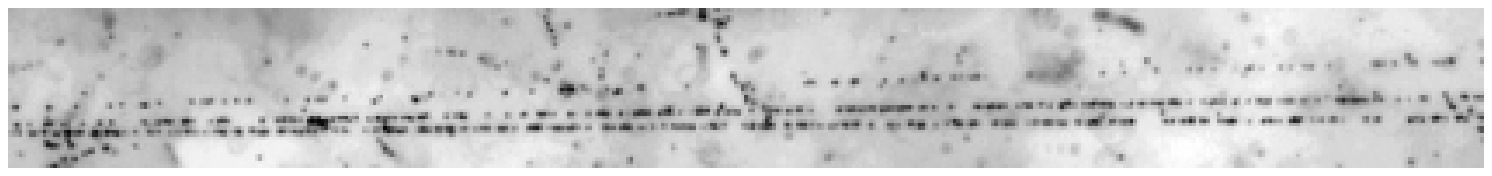}\\
  Shot~2 \\
 \end{tabular}}
\caption{\label{fig:2}Subsequently reconstructed photo of dissociation of a 1~A~GeV $^{10}$B nucleus into 1 
doubly and 2 singly charged fragments without production of target nucleus fragments and mesons (\lq\lq white\rq\rq star): 
interaction vertex (Shot~1); apparent tracks of 1 H and 2 He nuclei (from top to bottom, Shot~2).}
\end{figure*}
\indent The dissociation of the $^{10}$B nucleus at an energy of 1~A~GeV, studied in \cite{Adamovich04},
 may be an example which shows the dominance of the decay cluster channels. The microphotograph ~(see Fig.\ref{fig:2})~
 gives an event corresponding to the decay $^{10}$B$\rightarrow$2$^4$He$+$H. The fraction of
 these events is about 80\% of the total number of the events of the peripheral type.
 By measuring the multiple particle scattering, it is established that the deuterons
 participate in 40\% of the decay of such a type, in just the same way as in the $^6$Li nucleus
 with a pronounced $\alpha$-d structure. The decay $^{10}$B$\rightarrow^6$Li$+^4$He amounts
 to 15\% for a lower 4.5~MeV threshold. The decay $^{10}$B$\rightarrow^9$Be$+$p is only 3\% for a
 6~MeV threshold. Thus, for a giving type of the interaction of the $^{10}$B nucleus one has
 revealed the role of the 3-particle excitations as well as the role of the deuteron as
 a cluster element of the structure of this nucleus.\par
\begin{figure*}
\footnotesize
 \centerline{\begin{tabular}{@{}cc}
 \includegraphics[height=1.73 in]{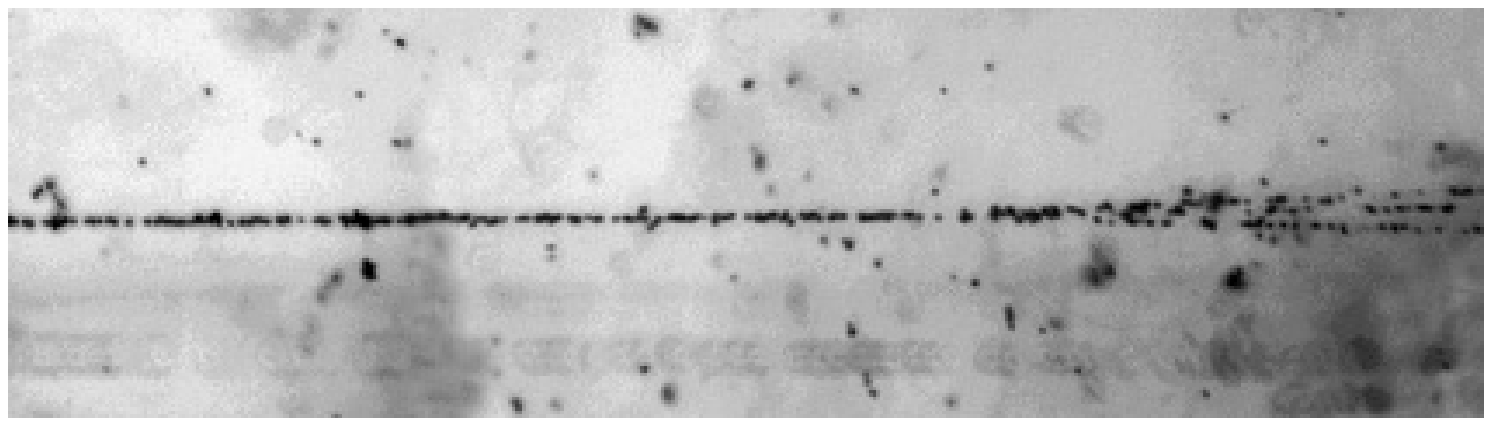}\\
  Shot~1  \\[8bp]
 \includegraphics[height=1.5 in]{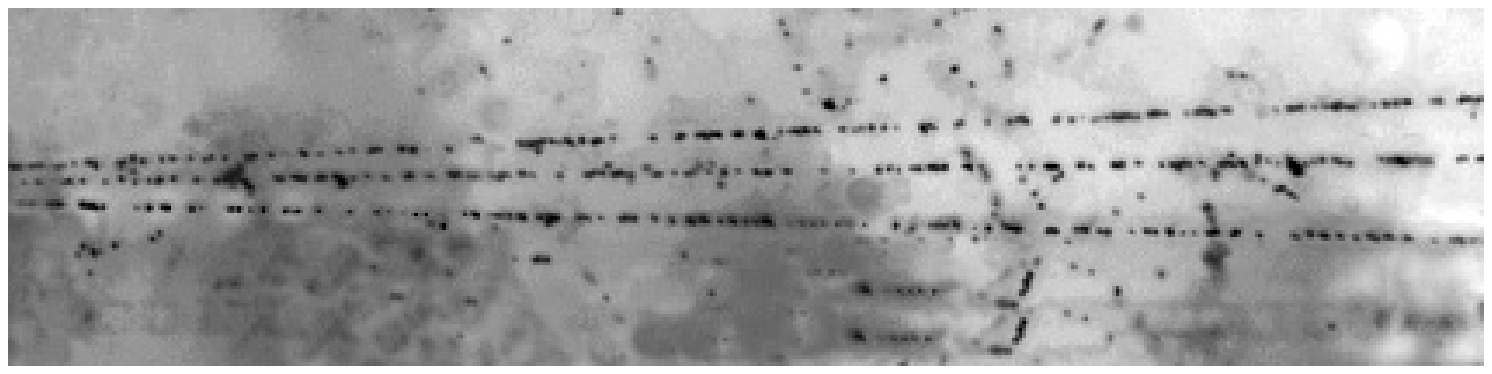}\\
  Shot~2  \\
 \end{tabular}}
\caption{\label{fig:3}Subsequently reconstructed photo of dissociation of a 1~A~GeV $^{10}$B nucleus into 3 
doubly charged fragments without production of target nucleus fragments and mesons (\lq\lq white\rq\rq star): 
interaction vertex (Shot~1); apparent tracks of 3 He nuclei (Shot~2).}
\end{figure*}

\indent The next microphotograph ~(see Fig.\ref{fig:3})~ shows the event of a three-particle decay of the $^{10}$B
 nucleus in the charge-exchange reaction without the production of a charged meson.
 Its charge topology may unambiguously be interpreted as $^{10}$B$\rightarrow$2$^3$He$+^4$He.
 Because of a deep rearrangement of nucleons which results in the formation a $^3$He cluster,
 an essentially higher (18~MeV) threshold should have been overcome in this event.
 Thus, there proceeds the population of a strongly excited state in a mirror $^{10}$C
 nucleus. This event points to the fact that in stellar media consisting of the $^{3,4}$He
 mixture there can occur an inverse process 
 which is similar to the 3$\alpha$ process 3$^4$He$\rightarrow$ $^{12}$C.
 The fusion process 2$^3$He$+^4$He$\rightarrow$ $^{10}$C results in a larger energy yield which is followed in the world of stable
 nuclei by the production of the $^{10}$B nucleus as a final product.\par
\indent This example illustrates the suggestion that the proof of the existence of the
 nuclear-physical process may become the basis for developing the ideas about the
 nucleosynthesis. However, in order to relate the relativistic fragmentation to thermonuclear
 synthesis it is necessary to establish an internal scale of kinetic energies in the c.m.s.
 of relativistic fragments.\par 
 \section{\label{sec:level5}The energy scale in N$\alpha$-particle systems}  
 \indent The study of the $^9$Be nucleus dissociation into two $\alpha$ particles allows one
 to restore their resonance states without a combinatorial background. Owing to a low-energy
 threshold (1.7~MeV) this process dominates the channel $^3$He$+^4$He$+$2n (22~MeV~threshold)
 which is similar to the latter by the image of the tracks. The separation of a neutron from
 the $^9$Be nucleus can lead to the production of an unstable $^8$Be nucleus with a decay
 through the ground state 0$^+$ (the decay energy is 92~keV, the width is 5.6~eV), as well as
 through the 1st 2$^+$ (3~MeV, 1.5~MeV width) and the 2nd 4$^+$ (11.4~MeV, 3.5~MeV width)
 excited states. On the basis of a reliable observation of theses states in the excitation
 spectrum Q, i. e. of the invariant mass of a pair of relativistic $\alpha$ particles minus
 their masses, it is possible to verify the validity of the excitation estimate using
 only the angular measurements.\par
\begin{figure}
\centerline{\includegraphics[width=3.5in]{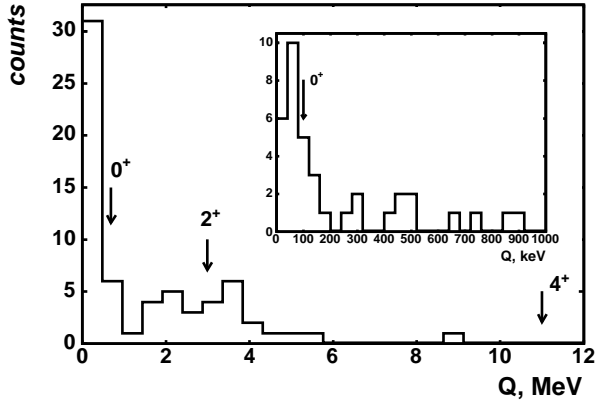}}
\caption{\label{fig:4}Distribution of $\alpha$ particle pairs $vs$ Q for the fragmentation mode $^9$Be$\rightarrow$2$\alpha$. 
Insertion: the distribution zoomed between 0-1000~keV.}%
\end{figure}

 \indent Emulsions are exposed to the secondary beam of $^9$Be nuclei of energy 1.2~GeV which
 is formed on the basis of the fragmentation of the $^{10}$B nuclei. At present, a total of
 160 stars with a pair of relativistic He nuclei have been found in the exposed material.
 The directions of their tracks are within the forward cone with about a 3$^{\circ}$ opening
 angle. The emission angles have already been measured for 70 events which allow one to present
 the spectrum Q of their excitation energies (see Fig.\ref{fig:4}).\par
 \indent The peaks correspond to the $^8$Be decay from the
 0$^+$ and 2$^+$. The scale of the spectrum part 0--1 MeV, given on the insertion of Fig.\ref{fig:4},
 is increased by a factor of 10. On it one can see a good coincidence of the distribution
 center with the decay energy of the $^8$Be ground state. The width of the
 peak makes it possible to define also the resolution of the method in this region of the
 spectrum. It is about 30~keV. On the right border of the distribution there are no events
 which would correspond to the second excited state 4$^+$ at 11.35~MeV energy. This spectrum
 area needs some additional statistics. Thus, in the system of two relativistic $\alpha$
 particles there are reflections of two known resonances.\par
\indent These excitations can be compared with those of more complicated N$\alpha$ particle
 systems. The comparison will be made for the events of the type of \lq\lq white\rq\rq stars,
 that is, the events which contain neither target-nucleus fragments, no produced mesons.
 In such events, there proceeds a more \lq\lq delicate\rq\rq excitation of the fragmenting
 nucleus. The excitation of the system is defined by the mean values of the transverse
 momenta of $\alpha$ particles. The $\alpha$ particle transverse momenta are calculated in
 the laboratory system by the formula P$_{T}$=4P$_{0}$sin$\theta$, where P$_{0}$ is the momentum
 per nucleon of a primary nucleus, $\theta$ the polar angle. As suggested in \cite{Belaga95}, the transition to the
 transverse momentum vectors in the c.m.s. is possible for small angles of scattering of the
 primary nucleus and is described by the following equation
 \begin{equation}
   {\bf P}^{*}_{Ti}\cong {\bf P}_{Ti}- \frac{\sum_{i=1}^{n}{\bf P}_{Ti}}{N_{\alpha}}\label{eq1} 
 \end{equation}
where, {\bf P}$_T$ is the residual vector of the $\alpha$ particle momenta.\par
 
\indent The mean values obtained for the alpha particle transverse momenta are as follows
 for $^9$Be$\rightarrow$2$\alpha$ $<$P$^*_T>$=103$\pm$2~MeV/c (the present paper),
 for $^{16}$O$\rightarrow$4$\alpha$ $<$P$^*_T>$=121$\pm$2~MeV/c $^{16}$O$\rightarrow$4$\alpha$ \cite{Andreeva96}, and for 
 $^{22}$Ne$\rightarrow$5$\alpha$ $<$P$^*_T>$=200$\pm$2~MeV/c (processing of the available
 data presented in \cite{El-Naghy88}). These values clearly demonstrate a tendency to an increase of the average
 momentum of the $\alpha$ particles with increasing their multiplicity and consequently an
 increase of the total Coulomb interaction of $\alpha$ clusters.\par
\begin{figure}
\centerline{\includegraphics[width=3.5in]{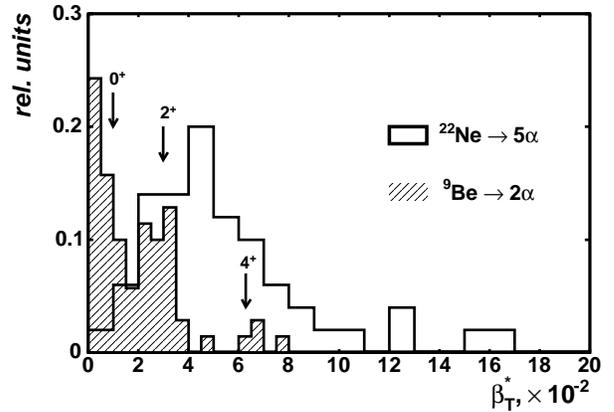}}
\caption{\label{fig:5}Distribution over the transverse velocities $\beta_T$ in
 the c.m.s. for the $\alpha$ particles produced in $^9$Be$\rightarrow$2$\alpha$ and
 $^{22}$Ne$\rightarrow$5$\alpha$ processes.}%
\end{figure}
\indent Fig.\ref{fig:5} shows the distribution over the transverse velocities $\beta_T$ in
 the c.m.s. for the $\alpha$ particles produced in $^9$Be$\rightarrow$2$\alpha$ and
 $^{22}$Ne$\rightarrow$5$\alpha$ processes. It can be concluded that the distribution over
 the velocities $\beta_T$, in spite of a broadening, is situated in a non-relativistic domain.
 In the $^9$Be case, the distribution has peaks which are due to the decay of the 0$^+$ and
 2$^+$ states. The distribution for the $^{22}$Ne nucleus is essentially broader, its mean
 value reflects the growth of the $\alpha$ particle transverse momenta, and the possible contribution
 from the decay of the $^8$Be 0$^+$ state is not significant. Thus, the energy
 calibration of the angular measurements by the $^9$Be nucleus enables one to conclude
 reliably about the features of internal velocity distributions in more complicated systems
 of relativistic $\alpha$ particles.\par
 \indent A nuclear state analogous to the dilute Bose gas can be revealed as the formation of N$\alpha$
 particle ensembles possessing quantum coherence near the production threshold.
 Being originated from relativistic nuclei it can appear in a form of narrow n$\alpha$ particle jets in the forward 
 cone. The predicted property of these systems is a narrow velocity distribution in the c. m. s.
 \cite{Horiuchi03}. The detection of
 such \lq\lq ultracold\rq\rq ~N$\alpha$ states is a serious argument
 in favor of the reality of the phase transition of $\alpha$ clusterized nuclei to the dilute Bose gas of $\alpha$
 particles. It gives a special motivation to explore lighter N$\alpha$ systems produced
 as potential \lq\lq building blocks\rq\rq of the dilute $\alpha$  particle Bose gas.\par
 \indent The behaviour of relativistic systems consisting of several H and He nuclei will be described
 in terms of invariant variables of a 4-velocity space as suggested in \cite{Baldin90}.  The relativistic projectile fragmentation results in the production of a
 fragment jet which can be defined by invariant variables characterizing relative motion
 \begin{equation}
    {b_{ik}=-\left( {P_i \over m_i} -{P_k \over m_k}\right)^2}
 \end{equation}
 with $P_{i(k)}$ and $m_{i(k)}$  being
 the 4-momenta and the masses of the $i$ or $k$ fragments.
 Following \cite{Baldin90}, one can suggest that a jet is composed of the nuclear fragments having relative motion within the non-relativistic range
 10$^{-4}$~$<$b$_{ik}<$~10$^{-2}$.
 The lower limit corresponds to the ground state decay $^{8}$Be$\rightarrow$2$\alpha$, while the
 upper one - to the boundary of low-energy nuclear interactions. The expression of the data via the
 relativistic invariant variable b$_{ik}$ makes it
 possible to compare the target and projectile fragmentation in a common form.
\section{\label{sec:level6}Conclusions}
\indent The degree of the dissociation of the relativistic nuclei in peripheral interactions
 can reach a total destruction into nucleons and singly and doubly charged fragments.
 In spite of the relativistic velocity of motion of the system of fragments as a whole,
 the relative motion of fragments is non-relativistic one. The invariant presentation makes it possible
 to extract qualitatively new information about few-cluster
 systems from fragmentation of relativistic nuclei in peripheral interactions. The emulsion technique allows one to
 observe these systems to the smallest details and gives the possibility of studying them
 experimentally\par
 \indent Investigations of the relativistic fragmentation of the nuclei ranging from Be
 to C can serve as some kind of \lq\lq bricks\rq\rq in the construction of a complete
 picture of the phase transition of heavy nuclei to the lightest clusters. In solving such
 problems, the nuclear energy of the order of several GeV per nucleon is optimal as far as at
 these energies the relativistic fragmentation cone has the optimal value which can be
 measured by microscopes.\par
 \indent  In the charge topology of the light nucleus fragments, the cluster character of
 their excitations is clearly manifested. The cluster degrees of freedom in nuclei are deeply
 associated with the process of their synthesis. The given event of the breakup with
 simultaneous charge-exchange of the $^{10}$B nucleus into 3 He nuclei points to a possible
 population of the excited states of the $^{10}$C nucleus with a deep rearrangement of the
 cluster structure of this nucleus. Observation of this process can point to the occurrence
 of an inverse fusion process 2$^3$He$+^4$He in stellar media.\par
 \indent For the particular case of the relativistic $^9$Be nucleus dissociation it is shown that precise
 angular measurements play a crucial role in the restoration of the excitation spectrum of
 the alpha particle fragments. This nucleus is dissociated practically totally through the 0$^+$
 and 2$^+$ states of the $^8$Be nucleus. \par
 \indent A detailed study of the nuclear fragment ensembles makes it possible to go on to
 the search for complicated quasi-stationary states of fragments. In the nuclear scale
 of distances and excitations they can possess properties which make them analogous to
 dilute quantum gases in atomic physics at ultra-cold temperatures. The proof of the
 existence of such systems can find some important applications for the problems of nuclear
 astrophysics. In this respect, the fragment jets are a microscopic model of stellar
 media.\par
\begin{acknowledgments}
In conclusion, we would like to remember  the names of our leaders in the domain of investigations 
with relativistic nuclei. Unfortunately, they are no more among the living. The foundations of 
the research  along these lines had been laid by Academician A.M.Baldin. For many years, M.I.Adamovich,
 V.I.Ostroumov, Z.I.Solovieva, K.D.Tolstov, M.I.Tretiakova, and G.M.Chernov had been leaders of the
 investigations  carried out by nuclear emulsion technique at the JINR synchrophasotron.
We hope that this sketch will be a contribution to a thankful memory of the senior generation scientists
in Russia.
\end{acknowledgments}

\end{document}